# Adaptive Super-twisting Second-order Sliding Mode for Attitude Control of Quadcopter UAVs

Van Truong Hoang[1], Quang Hieu Pham[2]
[1]University of Technology Sydney Broadway, New South Wales, Australia
[2]Naval Academy, Nha Trang, Khanh Hoa, Vietnam
E-mail: vanTruong.hoang@uts.edu.au, hieu.phamquang@gmail.com

## Abstract
This work addresses the modelling and control aspects for quadcopter or drone unmanned aerial vehicles (UAVs). First, the mathematical model of the drone is derived by identifying significant parameters and the negligible ones are treated as disturbances. The control design begins with the switching surface selection, then, an Adaptive Super Twisting Sliding Mode (ASTSM) Control algorithm is applied to adjust attitudes of the quadcopter under harsh conditions such as nonlinear, strong coupling, high uncertainties and disturbances. Simulation results show that the proposed controller can achieve robust operation with disturbance rejection, parametric variation adaptation as well as chattering attenuation. Comparisons with some commonly used and advanced controllers in a quadcopter model show advantages of the proposed control scheme.

## 1. Introduction

Quadcopters, known as quadrotors or drones, belong to a particular type of Vertical Take-Off and Landing aircraft with four directed rotors upward. The electric motors and their corresponding propellers are usually placed in a square formation with an equal distance to the centre of mass. Quadrotors are controlled by adjusting angular velocities of the propellers. They have been used in numerous real-world applications, such as surveillance, search and rescue operations, infrastructure inspections [1], emergencies management and product home delivery [2].

In research, the quadcopter is an exemplary design for small unmanned aerial vehicles with six degrees of freedom but only four independent inputs, thus, make it critically underactuated. To gain the six degrees of freedom, rotational and translational motions are coupled. As a result, dynamics of this flying object are highly nonlinear, particularly under the effect of the aerodynamics. Besides that, quadrotor has microscopic friction to prevent its movement, so it must yield its own damping to block the move and remain in a steady state. As a consequence, the design of controllers for the quadcopter becomes extreme problematic tasks.

A vast volume of controllers has been developed for quadrotors in literature, such as PID [3], H∞ [4], optimal [5], SMC [7], [8] and potential field [6]. Among them, SMC has been widely used because of its capability to robustly control systems under uncertainties and disturbances. Even though, chattering phenomenon remains as a significant disadvantage of the method. To eliminate chattering, high-order sliding modes (HOSM) [10], [11], [22] have been offered as a most likely preferable solution [12].

HOSM is a higher-order derivative of the conventional sliding mode for sliding function [10]. This creates an attraction for researchers to continuously develop related mathematical problems, accompanied by brilliant solutions, i.e., [12], [13]. HOSM is capable of removing the condition to have the relative degree to be equivalent to one for the tradition SMC and reduce the chattering effect. Another advantage of HOSM is in the construction of an accurate, robust differentiator with finite time convergence [14] or fixed-time convergence [15].

The second order sliding mode (SOSM) controllers, i.e., twisting, super twisting and accelerated twisting [9], [15], [16], [21], quasi-continuous [17], sub-optimal [18], and drift algorithm [19] have been extensively developed during the last two decades. The main idea of SOSM is not only to drive the sliding surface but also its derivatives to zero. Among them, super-twisting sliding mode (STSM) is a unique continuous sliding mode algorithm, which ensures all essential properties of the first-order SMC together with chattering rejection. However, the performance of STSM depends on the knowledge of the bound of perturbations. In practical scenarios, the drones are affected by disturbances, uncertainties, modelling errors and parameter variations that may downgrade the control efficiency, but their boundaries are not obvious. To address this concern, STSM controller with an adaptive gain has been applied to drive the switching variable and its derivative to zero in the presence of both additive and multiplicative disturbances [16].

In this paper, the ASTSM algorithm is proposed to control the attitude of quadcopters, which subject to nonlinear dynamics, strong coupling, high uncertainties and disturbances. The mathematical model of the drone is derived by adopting possible vital variables while some others are considered to be uncertainties. The controller mentioned above is proposed to achieve the robustness while rejecting disturbances and parametric variations as well as decreasing affection of the chattering phenomenon. This control performance is demonstrated by extensive simulation and comparison with the conventional Proportional-Integral-Derivative (PID), the classical first-order Sliding Mode (SMC) and the second-order Accelerated Twisting Sliding Mode (ATSM) controllers to show its advantageous feasibility.

The paper is organized as follow. The nonlinear dynamic model of quadrotor is presented in Section 2 followed by an introduction of second-order super twisting sliding mode with an adaptive scheme in Section 3. Section 4 describes simulation results and comparison strategies. The paper ends with a conclusion and recommendation for the future study.

## 2. Dynamic Model

### 2.1 Kinematics

The quadcopter model is shown in the Fig.1. Its dynamics is set up by two coordinate systems, namely earth frame (inertial frame) and the body fixed frame (body frame). The inertial frame $(x_E, y_E, z_E)$ is defined by the ground, with $z_E$ pointing down to the earth centre. The body frame $(x_B, y_B, z_B)$ is specified by the orientation of the quadcopter, with the rotor axes pointing downward and the arms pointing in $x_B$ and $y_B$ directions.

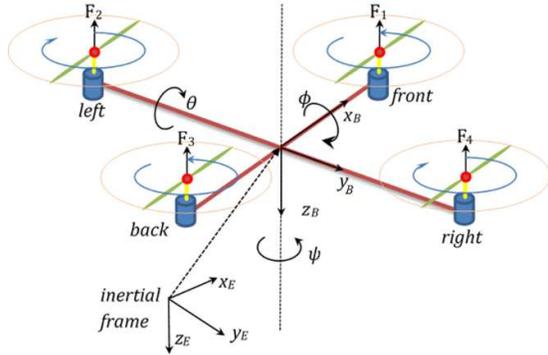

**Fig. 1** *A schematic diagram of quadcopter*

The equations representing the motion of the quadcopter are basically those of a rotating rigid body with six degrees of freedom, i.e., three translational and three rotational motions. The translational movements are defined in the earth frame, where the position is presented in vector form as $\xi = (x, y, z)^T$ and the vector $\dot{\xi} = (\dot{x}, \dot{y}, \dot{z})^T$ denotes its linear velocity. The drone attitude is defined by using the three Euler angles, named roll, pitch, and yaw are determined in the body frame as $\Theta = (\varphi, \theta, \psi)^T$, their corresponding angular rates are performed as $\dot{\Theta} = (\dot{\varphi}, \dot{\theta}, \dot{\psi})^T$.

Let $\omega = (p, q, r)^T$ represents the angular rate vector in the inertial frame. Then, the following rotational kinematics is achieved to show the relation between the earth angular velocity and the Euler angle rate vectors:

$$\omega = \begin{bmatrix} 1 & 0 & -s_\theta \\ 0 & c_\phi & c_\theta s_\phi \\ 0 & -s_\phi & c_\theta c_\phi \end{bmatrix} \quad (1)$$

where $s_x = sin(x)$ and $c_x = cos(x)$.

The below transformation matrix defines the relation between the body frame to earth frame translational velocities:

$$R = \begin{bmatrix} c_\psi c_\theta & c_\psi s_\theta c_\phi - s_\psi c_\phi & c_\psi s_\theta c_\phi + s_\psi s_\phi \\ s_\psi c_\theta & s_\psi s_\theta s_\phi + c_\psi c_\phi & s_\psi s_\theta c_\phi - c_\psi s_\phi \\ s_\theta & c_\psi s_\theta & c_\theta c_\phi \end{bmatrix} \quad (2)$$

### 2.2 Quadcopter Dynamics

Since the objective of this study is the attitude control, only torque elements that are capable of varying the quadcopter orientation are taken into account. They include torques caused by thrust forces $\tau$, body gyroscopic effects $\tau_b$, propeller gyroscopic effects $\tau_p$, and aerodynamic friction $\tau_a$.

The torque $\tau$ is produced by the quadcopter in the body frame including roll, pitch and yaw components, i.e., $\tau = [\tau_\phi, \tau_\theta, \tau_\psi]^T$. They are performed as:

$$\tau_\phi = l(F_2 - F_4) \quad (3)$$
$$\tau_\theta = l(-F_1 + F_3) \quad (4)$$
$$\tau_\psi = c(-F_1 + F_2 - F_3 + F_4) \quad (5)$$

where $F_k$, $k = 1...4$, is the thrust force generated by the propeller $k$, $l$ is the distance from a motor to the center of mass and $c$ is a force-to-torque scaling coefficient.

The body gyroscopic torque is modelled as:

$$\tau_b = S(\omega)I\omega \quad (6)$$

where $S(\omega)$ is a skew-symmetric matrix for the given vector $\omega$, and is expressed as follows:

$$S(\omega) = \begin{bmatrix} 0 & -r & q \\ r & 0 & -p \\ -q & p & 0 \end{bmatrix} \quad (7)$$

The resultant of torques generated by propeller gyroscopic effects $\tau_p$ is determined as:

$$\tau_p = \begin{bmatrix} I_r \Omega_r q \\ -I_r \Omega_r p \\ 0 \end{bmatrix} \quad (8)$$

where $I_r$ is the inertial moment of rotor, $\Omega_r = -\Omega_1 + \Omega_2 - \Omega_3 + \Omega_4$ is the residual angular velocity of rotor in which $\Omega_k$ denotes the angular velocity of the propeller $k$. The aerodynamic friction torque $\tau_a$ is given by:

$$\tau_a = k_a \omega^2 \quad (9)$$

where $k_a$ is a positive definite matrix of aerodynamic friction coefficients, $k_a = diag[k_{ax}, k_{ay}, k_{az}]$.

Using the aforementioned torques, the overall attitude dynamic model of the quadcopter is derived as:

$$I\ddot{\Theta} = \tau_b + \tau + \tau_p - \tau_a \quad (10)$$

where $I$ is a diagonal positive definite matrix of inertia tensors when the quadrotor is assumed to be symmetrical, $I = diag[I_{xx}, I_{yy}, I_{zz}]$.

In our study, the gyroscopic and aerodynamic torques are considered as external disturbances, and they are supposed to be removed by the advancement of the proposed controller. Therefore, the control inputs mainly depend on torque $\tau$ and from (3), (4) and (5), they can be represented as:

$$\begin{bmatrix} u_\phi \\ u_\theta \\ u_\psi \\ u_z \end{bmatrix} = \begin{bmatrix} \tau_\phi \\ \tau_\theta \\ \tau_\psi \\ F \end{bmatrix} = \begin{bmatrix} 0 & l & 0 & -l \\ -l & 0 & l & 0 \\ -c & c & -c & c \\ 1 & 1 & 1 & 1 \end{bmatrix} \begin{bmatrix} F_1 \\ F_2 \\ F_3 \\ F_4 \end{bmatrix} \quad (11)$$

where $u_\phi, u_\theta$ and $u_\psi$ respectively represent the roll, pitch and yaw torques, $u_Z$ represents the total thrust acting on the four rotors and $F$ denotes the UAV lift produced by the four propellers, $F = \sum_{i=1}^{k} F_k$. In this paper, $u_Z$ is supposed to accommodate with the gravity when we consider the rotational control only. In view of the equations from (3) to (7), the second-order nonlinear dynamics of quadcopters for attitude control can be described by the following equations:

$$\ddot{\phi} = \frac{I_{yy} - I_{zz}}{I_{xx}} qr + \frac{1}{I_{xx}} u_\phi + \frac{1}{I_{xx}} d_\phi \quad (12)$$

$$\ddot{\theta} = \frac{I_{zz} - I_{xx}}{I_{yy}} pr + \frac{1}{I_{yy}} u_\theta + \frac{1}{I_{yy}} d_\theta \quad (13)$$

$$\ddot{\psi} = \frac{I_{xx} - I_{yy}}{I_{zz}} pq + \frac{1}{I_{zz}} u_\psi + \frac{1}{I_{zz}} d_\psi \quad (14)$$

where $d_\varphi, d_\theta$ and $d_\psi$ are the disturbances, including the two terms $\tau_a$ in (8), $\tau_p$ in (9) to the system in real time. Let us define the following state variables:

$$\begin{cases} X_1 = \Theta \\ X_2 = \dot{\Theta} \end{cases} \quad (15)$$

Then, the dynamics of quadcopters can be represented in the following form as:

$$\begin{cases} \dot{X}_1 = X_2 \\ \dot{X}_2 = I^{-1}[f(X) + u(t) + d(t)] \end{cases} \quad (16)$$

where $u = [u_\phi, u_\theta, u_\psi]^T$ is the input vector, $d$ is the disturbance vector, $d = [d_\phi, d_\theta, d_\psi]^T$, and $f(X)$ is represented as:

$$f(X) = -S(\omega)I\omega = \begin{bmatrix} (I_{yy} - I_{zz})qr \\ (I_{zz} - I_{xx})pr \\ (I_{xx} - I_{yy})pq \end{bmatrix} \quad (17)$$

In our system, the following assumptions are made:
**A.1** The quadcopter structure is rigid and symmetric. The propellers are rigid.
**A.2** The signals $\Theta$ and $\dot{\Theta}$ can be measured by on-board sensors.
**A.3** The reference trajectories and their first and second time derivatives are bounded.
**A.4** The velocity and the acceleration of the quadcopter are bounded.
**A.5.** The orientation angles are limited to $\phi \in \left[-\frac{\pi}{2}, \frac{\pi}{2}\right]$, $\theta \in \left[-\frac{\pi}{2}, \frac{\pi}{2}\right]$ and $\psi \in [-\pi, \pi]$
**A.6** The rotational speeds of rotors are bounded.

## 3. Control Design

The control signals $u_\phi, u_\theta$ and $u_\psi$ in (16) are used to control the Euler angle $\Theta = [\varphi, \theta, \psi]^T$ to reach the reference value $\Theta_d = (\varphi_d, \theta_d, \psi_d)^T$.

The overall control law is presented as:

$$u(t) = u_{eq}(t) + u_D(t) \quad (18)$$

where $u_{eq}(t)$ is a continuous part defined by the controlled variables and reference values, $u_D(t)$ is the discontinuous part that contains a switching element.

### 3.1. Sliding Manifold

The sliding surface equation determines the dynamics of the system, so it is presented as:

$$\sigma = \dot{e} + \Lambda e \quad (19)$$

where $\Lambda = diag(\lambda_\varphi, \lambda_\theta, \lambda_\psi)$ is a positive definite matrix to be designed, and $e$ is the control error:

$$e = X_1 - X_{1d}$$

where $X_{1d}$ is the vector of desired angles. Thus, the first derivative of the error vector will be:

$$\dot{e} = \dot{X}_1 - \dot{X}_{1d}$$

### 3.2. Design $u_{eq}$

The equation (19) can be rewritten for the quadcopter attitude sliding surface as:

$$\sigma = (\dot{X}_1 - \dot{X}_{1d}) + \Lambda(X_1 - X_{1d}) \quad (20)$$

Taking time derivative of $\sigma$ we have:

$$\dot{\sigma} = (\ddot{X}_1 - \ddot{X}_{1d}) + \Lambda(\dot{X}_1 - \dot{X}_{1d}) \quad (21)$$

or

$$\dot{\sigma} = -\ddot{X}_{1d} + \dot{X}_2 + \Lambda \dot{e} \quad (22)$$

When the system is in its nominal condition, i.e., $d(t) = 0$, we can substitute $\ddot{X}$ from (16) into (22), which yields:

$$\dot{\sigma} = -\ddot{X}_{1d} + I^{-1}[f(X) + u + d] + \Lambda \dot{e} \quad (23)$$

During the time the system is in sliding phase, $u$ can be considered as the equivalent control $u_{eq}$. By driving the derivative of sliding surface to zero, we found the control rule for the continuous part:

$$u_{eq} = I(\ddot{X}_{1d} - \Lambda \dot{e}) - f(X) \quad (24)$$

### 3.3. Design $u_D$ and problem formulation

The second-order super twisting sliding mode controller is given in [10], $u_D$ is redefined as follows:

$$\begin{cases} u_D = -\alpha\sqrt{|\sigma|}\,sign(\sigma) + \nu \\ \dot{\nu} = -\dfrac{\beta}{2}sign(\sigma) \end{cases} \quad (25)$$

Where $\alpha$ and $\beta$ are definite positive diagonal matrices of corresponding gains to be adjusted.

Thus, we have the accomplished control equation for the quadrotor attitude:

$$u = I(\ddot{X}_{1d} - \Lambda\dot{e}) - f(X) + u_D \quad (26)$$

with $f(X) = -S(\omega)I\omega$, we can represent (26) in the following form:

$$u = I(\ddot{X}_{1d} - \Lambda\dot{e}) + S(\omega)I\omega + u_D \quad (27)$$

The quadcopter uncertainties are subjected to variations, modelling errors, as well as some disturbances such as aerodynamic frictions, propeller gyroscopic effects and environmental affections, particularly wind gusts while flying outdoor. Let $I = I_0 + \Delta I$ where $I_0$ and $\Delta I$ represent the known nominal and unknown uncertain parts of the inertia matrix, respectively. The term $\dot{X}_2$ in (16) becomes:

$$\dot{X}_2 = (I_0 + \Delta I)^{-1}S(\omega)I_0\omega + (I_0 + \Delta I)^{-1}u + (I_0 + \Delta I)^{-1}d + (I_0 + \Delta I)^{-1}S(\omega)\Delta I\omega \quad (28)$$

We have

$$(I_0 + \Delta I)^{-1} = I_0^{-1}(1 + I_0^{-1}\Delta I)^{-1} = AI_0^{-1} + B(1 + I_0^{-1}\Delta I)^{-1} \quad (29)$$

With $A$ and $B$ are diagonal matrices of constants found by breakdown analysis (29).

The sliding surface (23) will be derived as:

$$\dot{\sigma} = -\ddot{X}_{1d} + \Lambda\dot{e} + (I_0 + \Delta I)^{-1}S(\omega)I_0\omega + (I_0 + \Delta I)^{-1}u + (I_0 + \Delta I)^{-1}d + (I_0 + \Delta I)^{-1}S(\omega)\Delta I\omega \quad (30)$$

Substitute (29) to (30), we have:

$$\dot{\sigma} = -\ddot{X}_{1d} + \Lambda\dot{e} + AS(\omega)I_0\omega + (I_0 + \Delta I)^{-1}S(\omega)\Delta I\omega + B(1 + I_0^{-1}\Delta I)^{-1}S(\omega)I_0\omega + (I_0 + \Delta I)^{-1}d + [AI_0^{-1} + B(1 + I_0^{-1}\Delta I)^{-1}]u \quad (31)$$

Let we rewrite $\dot{\sigma}$ in the following form:

$$\dot{\sigma} = a(x,t) + b(x,t)u \quad (32)$$

Where the function $a(x,t) \in \mathbb{R}^3$ is presented as:

$$a(x,t) = a_1(x,t) + a_2(x,t),$$
$$a_1(x,t) = -\ddot{X}_{1d} + \Lambda\dot{e} + AS(\omega)I_0\omega$$
$$a_2(x,t) = (I_0 + \Delta I)^{-1}S(\omega)\Delta I\omega + B(1 + I_0^{-1}\Delta I)^{-1}S(\omega)I_0\omega + (I_0 + \Delta I)^{-1}d$$

We assume that $a_1(x,t)$ and $a_2(x,t)$ bounded, but their boundaries are not yet clarified. Also, the function $b(x,t) \in \mathbb{R}^3$ is uncertain and represented as:

$$b(x,t) = b_0(x,t) + \Delta b(x,t),$$

where $b_0(x,t) = AI_0^{-1}$ and $\Delta b(x,t) = B(1 + I_0^{-1}\Delta I)^{-1}$ are a known function and a bounded perturbation, respectively. An assumption for this case is:

$$\Delta b(x,t)/b(x,t) < \gamma(x,t) < \gamma_1 < 1,$$

with $\gamma_1$ is an unknown boundary. Thus, it can be seen clearly that the input-output dynamics (32) contains of both additive and multiplicative perturbations.

The STSM controller (25) can robustly handle the given problems while their boundaries are known. However, the bound is not easy to evaluate in practice and besides, a high value of sliding gain $\alpha$ and/or $\beta$ will lead to high chattering magnitude. Therefore, the problem is now to drive the sliding variable $\sigma$ and its derivative $\dot{\sigma}$ to zero in finite time by means of super-twisting SMC without exaggeration of the control gains.

### 3.4. Adaptive STSM Control Design

The adaptive gains for (25) is defined as:

$$\alpha = \alpha(\sigma,\dot{\sigma},t) \quad (33a)$$
$$\beta = \beta(\sigma,\dot{\sigma},t) \quad (33b)$$

The ASTSM control gains [16] are proposed to decrease the chattering phenomenon and converge $\sigma$ and $\dot{\sigma}$ to zero in the presence of disturbances and uncertainties. The gains are chosen as:

$$\dot{\alpha} = \begin{cases} \varpi\sqrt{\dfrac{\gamma_1}{2}}sign(|\sigma|-\mu) & if\ \alpha > \alpha_m \\ \eta & if\ \alpha \leq \alpha_m \end{cases} \quad (34)$$

$$\beta = 2\varepsilon\alpha \quad (35)$$

where $\varpi > 0, \gamma_1, \mu, \eta$ and $\varepsilon$ are arbitrary positive scalars, $\alpha_m$ is the threshold of the adaptation. The significant property of the adaptive scheme is non-exaggerating the values of the gains to be adjusted. The global Lyapunov function candidate is defined as follows:

$$V = V_0(\sigma,\dot{\sigma}) + \dfrac{1}{\gamma_1}(\alpha - \alpha^*)\dot{\alpha} + \dfrac{1}{\gamma_2}(\beta - \beta^*)\dot{\beta} \quad (36)$$

Where $V_0(\sigma,\dot{\sigma})$ is a Lyapunov function for $(\sigma,\dot{\sigma})$, $\gamma_1$ and $\gamma_2$ are arbitrary positive numbers, $\alpha^*$ and $\beta^*$ are maximum possible values of $\alpha$ and $\beta$.

The derivative of the Lyapunov function (36) is obtained as:

$$\dot{V}(\sigma,\alpha,\beta) \leq -\eta_0\sqrt{V(\sigma,\alpha,\beta)} + \xi \quad (37)$$

where $\eta_0$ is a positive constant, $V(\sigma,\alpha,\beta) \geq 0$ is a function of $\sigma, \alpha$ and $\beta$, and

$$\xi = -|\varepsilon_\alpha|\left(\dfrac{1}{\gamma_1}\dot{\alpha} - \dfrac{\omega_1}{\sqrt{2\gamma_1}}\right) - |\varepsilon_\beta|\left(\dfrac{1}{\gamma_2}\dot{\beta} - \dfrac{\omega_2}{\sqrt{2\gamma_2}}\right) \quad (38)$$

where $\varepsilon_\alpha = \alpha - \alpha^*, \varepsilon_\beta = \beta - \beta^*$, $\omega_1$ and $\omega_2$ are some arbitrary positive constants.

It can be seen that the finite time convergence of the system is guaranteed given (34) and (35) [8].

## 4. Simulation Results

This section presents extensive simulations to demonstrate the performance of the ASTSM controller. The quadcopter model is the 3DR Solo drone shown in Fig. 2. Technical specifications and accessories of the quadcopter are described in [20]. The physical parameters $L(\cdot), d(\cdot), r(\cdot)$ and $h(\cdot)$ in the figure are measured distances, which are used to calculate model properties, as listed in Table I. The parameters of the proposed controller, are shown in Table 2.

Numerical simulation results have been done in three different conditions, i.e., responses of the system in nominal conditions, under the appearance of disturbances and parametric variations. The initial conditions of the quadrotor are assumed to be in its steady state, where all control angles and angular velocities are zeros. The desired angles are changed in the simulation as $\phi = -10^0$, $\theta = 10^0$ and $\psi = 45^0$ at time 0.5s, 1s and 2s, respectively.

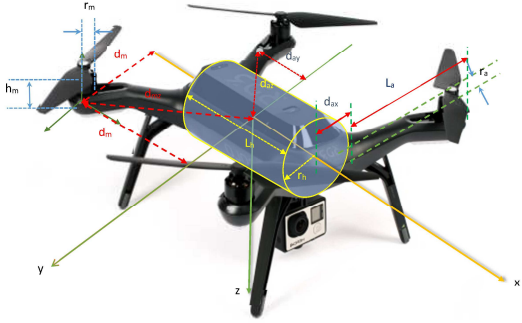

**Fig. 2** *The 3DR Solo drone with body coordinate frame.*

In disturbance scenario, a torque of 0.5N.m is separately added in each axis of the drone. Particularly, to demonstrate the performance of the ASTSM controller in dynamic variation conditions, simulation parameters are varied to counteract some modelling errors, the most capable payload 0.8 kg of the 3DR Solo, is added to the model and the inertial matrix is varied with the uncertainties as in Table 3.

*Table 1. Parameters of the quadcopter model*

| Parameter | Value | Unit |
|---|---|---|
| $m$ | 1.50 | $kg$ |
| $l$ | 0.205 | $m$ |
| $g$ | 9.81 | $m/s^2$ |
| $I_{xx}$ | $8.85 \cdot 10^{-3}$ | $kg.m^2$ |
| $I_{yy}$ | $1.55 \cdot 10^{-3}$ | $kg.m^2$ |
| $I_{zz}$ | $23.09 \cdot 10^{-3}$ | $kg.m^2$ |

*Table 2. Control design parameters*

| Variable | Value | Variable | Value |
|---|---|---|---|
| $\lambda_\varphi$ | 3.89 | $\varpi$ | 200 |
| $\lambda_\theta$ | 3.89 | $\gamma_1$ | 6.60 |
| $\lambda_\psi$ | 4.36 | $\varepsilon$ | 0.60 |
| $\eta$ | 0.01 | $\alpha_m$ | 0.01 |

*Table 3. Uncertainties added to the inertia matrix*

| $\Delta I$ | x | y | z |
|---|---|---|---|
| x | 0.4825 | 0.0044 | -0.0077 |
| y | 0.0044 | 0.2437 | 0.0115 |
| z | -0.0077 | 0.0115 | 0.2437 |

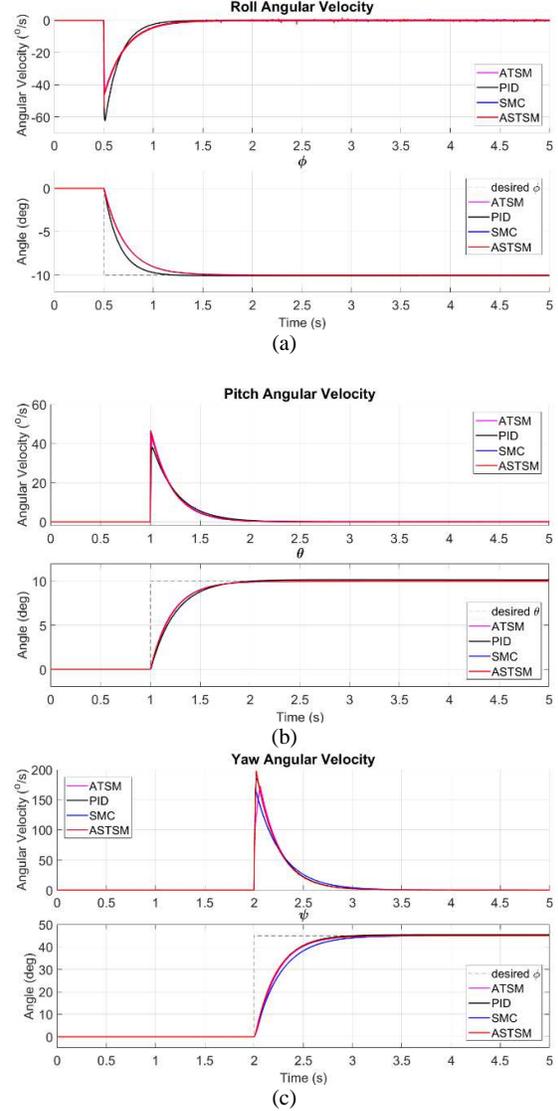

**Fig. 3** *Responses of angles and their angular velocities in nominal condition*

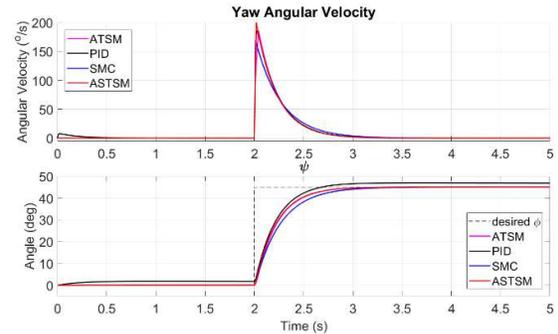

**Fig. 4** *Yaw angle and angular velocity responses in the presence of external disturbances*

Results of simulation are additionally compared with the ATSM [15], the conventional SMC, and the PID controller that is practically implemented to the Solo drone. Results of tracking behaviour in nominal situations are shown in the Fig.3. The outputs of the controllers shown in Fig.5, where the time scale is zoomed in to observe the gain response to adapt to various changes of the system, such as references and coupling effects. The adaptive gain of the ASTSM controller for yaw ($\alpha_3(t)$) responses is shown in Fig.7. The system response of the distinctive controllers under disturbances are presented in Fig.4. Fig.6 shows the simulation results in parametric variation in comparison with the nominal conditions.

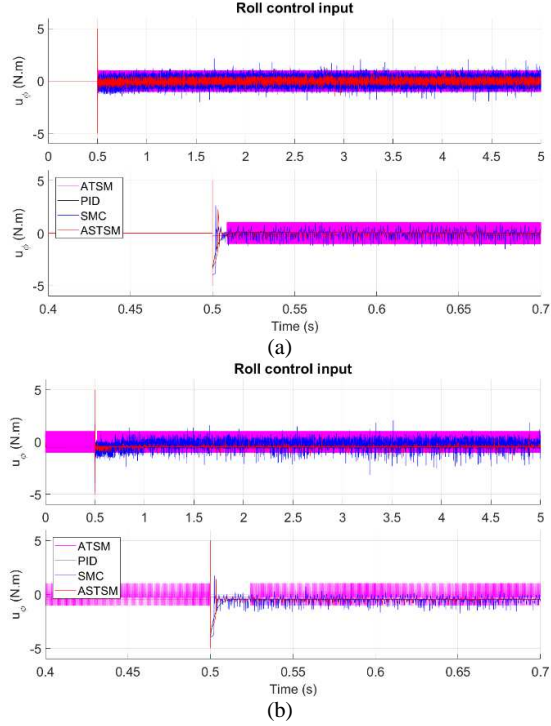

**Fig. 5** *Roll control inputs ($u_1$) in: (a) Nominal condition; and (b) Occurrence of disturbances*

Fig.3 illustrates that the controllers smoothly drives the angles to the required values with comparatively unremarkable overshoot within two seconds for all cases mentioned above. Robustness of the three sliding controllers under the presence of disturbances is presented in Fig.4. However, by the advantage of its adaptation, the designed controller performs its faster convergence.

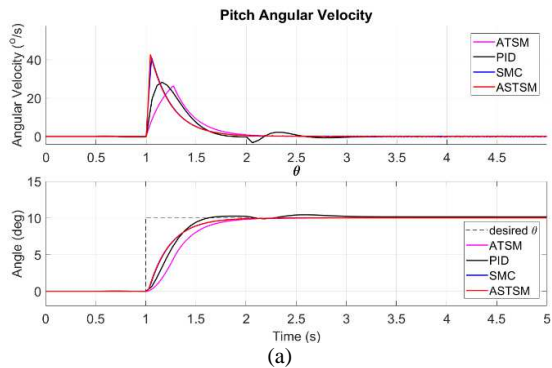

(a)

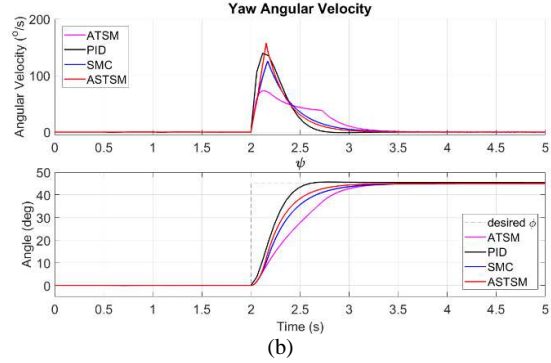

(b)

**Fig. 6** *Pitch (a) and yaw (b) angle and angular velocity responses in the presence of parametric variations*

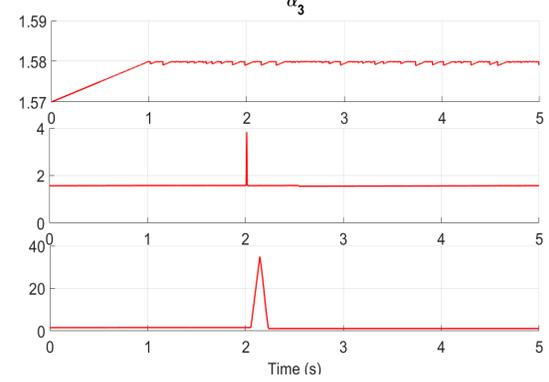

**Fig. 7** *The adaptation of gain $\alpha_3(t)$ in three scenarios: Top: Nominal condition; Middle: Occurrence of disturbances; Bottom: Parametric variations.*

The leading cause of chattering effect is modelling errors in conjunction with high gain selected to preserve the robustness of the system under perturbations [12]. It results in high chattering amplitude and frequency at the origin, as shown in case of ATSM and SMC control inputs in Fig.5. By comparing control inputs in the roll attitude, chattering effect is attenuated greater by ASTSM when the adaptive gain is adjusted to its threshold. Furthermore, the smaller magnitude of the control inputs compared to ATSM, SMC and PID illustrate that less energy is required by the designed control scheme.

There exist strong coupling relations among the control variables as pointed out in Eqs. (12-14) and the response of PID controller in Fig.6a shows clearly this phenomenon. However, the improvement of the proposed controller is able to solve this issue by managing the Euler angles to reach their desired values and then track them with no relative perturbation.

Responses of the system with parametric variations in Fig.6 show that ASTSM is capable of preserving its robust properties better than ATSM, SMC and PID in particular. It can be seen that the three sliding controllers reach the reference value without causing much overshoot or oscillation. However, the faster responses of ASTSM in the two above cases, again, illustrates its advancement.

Time histories of $\alpha_3(t)$ in Fig.7 show how the adaptive gain adjusted to response with the sudden change of the desired signals, coupling effect, disturbances (Fig.5a) and variation (Fig.5b). The higher gain magnitudes are observed in the two bottom subfigures imply more energy is required to stabilise the system in dealing with disturbances and uncertainties. This also suggests the feasibility of the control scheme.

## 5. Conclusion and Future work

The paper deals with the design of the adaptive second order sliding mode controller for a practical quadrotor. It begins with an introduction of refinements of nonlinear dynamic equations for the drone. An adaptive super-twisting sliding mode controller has been implemented to ensure robustness with respect to unknown bounded disturbances and uncertainties. The control design is based on the selection of a sliding surface and some parameters for adaptation of the control gain taking account into chattering reduction. Control performance is evaluated in simulation for the cases of both external disturbances and system uncertainties. The validity of the proposed control scheme is also judged through comparison with the accelerated twisting sliding mode, the conventional sliding mode and PID controllers. Future plan will focus on implementing the proposed controller to demonstrate the feasibility of the proposed approaches.